How physics instruction impacts students' beliefs about learning physics: A meta-analysis of 24 studies

By Adrian Madsen[1], Sarah B. McKagan[1], and Eleanor C. Sayre[2]
[1]American Association of Physics Teachers, College Park, MD
[2]Kansas State University, Manhattan, KS

**Abstract**

In this meta-analysis, we synthesize the results of 24 studies using the Colorado Learning Attitudes about Science Survey (CLASS) and the Maryland Physics Expectations Survey (MPEX) to answer several questions: (1) How does physics instruction impact students' beliefs? (2) When do physics majors develop expert-like beliefs? and (3) How do students' beliefs impact their learning of physics? We report that in typical physics classes, students' beliefs deteriorate or at best stay the same. There are a few types of interventions, including an explicit focus on model-building and/or developing expert-like beliefs that lead to significant improvements in beliefs. Further, small courses and those for elementary education and non-science majors also result in improved beliefs. However, because the available data oversamples certain types of classes, it is unclear whether these improvements are actually due to the interventions, or due to the small class size, or student population typical of the kinds of classes in which these interventions are most often used. Physics majors tend to enter their undergraduate education with more expert-like beliefs than non-majors and these beliefs remain relatively stable throughout their undergraduate careers. Thus, typical physics courses appear to be selecting students who already have strong beliefs, rather than supporting students in developing strong beliefs. There is a small correlation between students' incoming beliefs about physics and their gains on conceptual mechanics surveys. This suggests that students with more expert-like incoming beliefs may learn more in their physics courses, but this finding should be further explored and replicated. Some unanswered questions remain. To answer these questions, we advocate several specific types of future studies: measuring students' beliefs in courses with a wider range of class sizes, student populations, and teaching methods, especially large classes with very innovative pedagogy and small classes with more typical pedagogy; analysis of the relationship between students' beliefs and conceptual understanding including a wide variety of variables that might influence each; and analysis of large data sets from a variety of classes that track individual students rather than averaging over classes.

## I. INTRODUCTION AND SUMMARY OF FINDINGS

Physics faculty care about their students learning physics content. In addition, they usually hope that their students will learn some deeper lessons about thinking critically and scientifically. They hope that as a result of taking a physics class, students will come to appreciate physics as a coherent and logical method of understanding the world, and recognize that they can use reason and experimentation to figure things out about the world. While it is relatively straightforward to measure students' understanding of physics content, it is much more difficult to measure how much they think like a physicist and what they believe about the nature of physics and learning physics. Physics education researchers have created several surveys to measure students' beliefs about physics[1–5].

Two commonly used surveys are the Maryland Physics Expectations Survey (MPEX)[1] and the Colorado Learning Attitudes about Science Survey (CLASS)[2]. These surveys are not about whether students like physics, but about how students perceive the discipline of physics or their particular physics course. These surveys ask students to rank statements using a 5-point Likert scale from strongly agree to strongly disagree. The most common way to score these surveys is to collapse students' responses into a binary depending on whether they are the same as an expert physicist would give (we will call this "percent expert-like response", in the literature it is commonly called "percent favorable response"). They are usually given as a pre- and posttest to measure the "shift", or change in students' beliefs over the course of a semester. Work has also been done using these surveys to study how students' beliefs change over the course of their undergraduate careers.

These surveys have been given to thousands of students at a variety of institutions across the US. A rich set of results has emerged on how the scores vary by teaching method, over time and correlate with other measures (such as the Force Concept Inventory (FCI)[6] and the Force and Motion Conceptual Evaluation (FMCE)[7]). In this paper, we bring together these results to answer several questions:

1. How does physics instruction impact students' beliefs?
2. When do physics majors develop expert-like beliefs?
3. How do students' beliefs impact their learning of physics?

## II. HOW CAN WE ASSESS STUDENTS' BELIEFS ABOUT LEARNING PHYSICS?

To answer our questions about how physics instruction impacts students' beliefs about learning physics, we examine the set of published results on the CLASS and MPEX. These surveys measure students' self-reported beliefs about physics and how closely these beliefs about physics align with experts' beliefs. The surveys ask students questions about how they learn physics, how physics is related to their everyday lives, and how they think about the discipline of physics. Example questions are given in Table 1. A list of the categories of questions for each test is available in Table 11 in the Appendix. Because this is self-reported data, we can't know how well the beliefs students report on the CLASS or MPEX correspond to the ways they actually think about physics. For example, a student might say and really believe, "When I am solving a physics problem, I try to decide what would be a reasonable value for the answer," but not do that in real life.

We picked the CLASS and MPEX because they are the most commonly used surveys of beliefs about physics. Further, the questions on these two tests overlap to some degree and they are designed to measure a similar aspect of students' beliefs, so it seems reasonable to look at the results of these tests together. Table 11 in the Appendix describes several important characteristics of each survey.

**Table 1.** Example items from the CLASS and MPEX. Statements that experts agree with are indicated in bold.

| CLASS | MPEX |
| --- | --- |
| • **I study physics to learn knowledge that will be useful in my life outside** | • **Learning physics helps me understand situations in my** |
| **of school.** | **everyday life.** |
| • A significant problem in learning physics is being able to memorize all the information I need to know. | • A significant problem in this course is being able to memorize all the information I need to know. |
| • Knowledge in physics consists of many pieces of information each of which applies primarily to a specific situation. | • Knowledge in physics consists of many disconnected topics. |
| • **If I get stuck on a physics problem on my first try, I usually try to figure out a different way that works.** | • In doing a physics problem, if my calculation gives a result that differs significantly from what I expect, I'd have to trust the calculation. |

## III. HOW DOES PHYSICS INSTRUCTION IMPACT STUDENTS' BELIEFS?

To determine how physics instruction impacts students' beliefs, we conducted a literature review to find relevant studies. We identified 24 studies published in *Physical Review Special Topics – Physics Education Research*, *American Journal of Physics*, and the *Physics Education Research Conference Proceedings* that reported CLASS or MPEX results for undergraduate physics classes in the US and Canada. Because of the variability of the education systems in other countries, we did not include studies from outside the US and Canada. For those interested in these international studies, see references[8–11]. Where there was insufficient information, such as missing information about error or average pre- and posttest scores, we contacted the authors for clarification and/or additional data. There were three studies where the authors did not write back or the error information was no longer available. There were an additional three studies where the authors did not calculate error.

We chose to look at the shifts in overall average percent expert-like responses (shifts) from pre- to posttest on the CLASS or MPEX. This shift is found by:

(1) determining, for each student, the percentage of the responses that are in agreement with the expert answer on the pre- and posttest
(2) averaging these individual scores to find the average percent expert-like on the pre- and posttest for the entire class
(3) subtracting the pretest average percent expert-like from the posttest average percent expert-like.

This metric tells us how students' expert-like beliefs about physics changed from the start to end of their physics course. Ideally, we would like this value to increase as a result of physics instruction. We chose this metric instead of average percent unfavorable scores because it is more commonly reported in the literature. Of the 24 studies we examined, only seven reported the shift in percent unfavorable scores [1,4,12–15]. Further, the shifts in favorable beliefs (expert-like beliefs) are almost always in the opposite direction to the shifts in unfavorable beliefs (novice-like beliefs), as expected, so looking only at favorable shifts gives us similar information as looking at unfavorable beliefs. Of the seven studies which reported unfavorable beliefs, only one course[14] reported positive shifts in both the favorable and unfavorable scores (this can happen when students shift from neutral answer choices on the pretest to positive and negative answer choices on the posttest).

Studies that used matched data, those where shifts are calculated using the same individual students for both the pre- and posttest, as opposed to using the class average pretest and posttest scores, most accurately represent the change in students' beliefs. This is because shifts calculated using unmatched data are more susceptible to selection effects, as students with less expert-like beliefs are most likely to drop the course or not fill out the survey at the end of the semester. Of the 24 studies included in our analysis, 12 reported using matched data. Four studies reported using matched conceptual assessment data, but did not report whether their CLASS or MPEX data were matched. Six studies did not report whether their data were matched or not, though five of these six studies gave one "n" value for the number of students in their dataset. One of these six studies reported only the number of students enrolled in the course. Based on this information, we can assume that most of the studies included in our analysis use matched data, but we can't know for sure.

The questions on both the MPEX and CLASS can be clustered into different categories. We chose to focus on the overall shift (for all categories) instead of looking at individual categories because this is what is consistently reported in almost all studies. Further, the categories on the MPEX and CLASS are different, making comparisons between the two difficult.

Figure 1 shows the shifts grouped by their direction (negative[1,2,14–20], positive[4,12,13,21–31], or no shift[1,14–16,18,19,21,32]). The direction of the shift is that which was reported in the study. We report the number of students for each study as given in the paper. Where there was more than one course at the same institution taught using the same teaching method, we used a weighted average to combine the results. We were not able to obtain error for several studies, so we were not able to determine effect size. Our meta-analysis includes results from both the CLASS and MPEX. In Figure 1, we see that the shifts for MPEX data are comparable to the shifts for CLASS data for similar types of courses and institutions, so we believe that further normalization of the results from these different tests is not necessary. There is one positive shift bar in Figure 1, "Physics by Inquiry, Research University [21]", where the error bars overlap zero. In this study, the distributions of CLASS scores deviated from the normal distribution, so statistical significance of the gains was determined using the nonparametric Wilcoxon signed-rank test. When using the Wilcoxon signed-rank test, error bars can overlap zero, and the result can still be statistically significant. See Krzywinski and Altman[33] for a discussion of this test and an illustration of error bars which overlap zero but are still significant at the 0.04 level.

Using Figure 1, we looked for similarities between the courses in each shift category. We noticed many factors that may have contributed to the differences in the shift categories including teaching method, class size, population, and pretest scores. Many of these factors are confounded; for example, most of the classes with large positive shifts are those using curricula with a focus on model building, which are implemented in small classes and often offered to future elementary teachers. Courses with negative shifts tend to be calculus-based with large enrollments and taught in a reformed or traditional manner.

We conducted a three-way type I ANOVA to test the influence of teaching method, class size, and student population on the CLASS/MPEX shifts. We chose a type I ANOVA because there is an inherent ordering of these factors according to what is more educationally easy to change. Student population is ordered first in our analysis as this is the least easy for a faculty member to change. Next, class size is widely determined outside of an instructor's purview. Third in order is teaching method, as this is something a faculty member could indeed have influence over.

We found no significant three-way interaction or two-way interactions in our analysis. We tested the main effects of each factor. All three factors were found to be significant at the $p < .01$ level (Table 2). We used a Tukey HSD post-hoc test to determine which levels of the factors were significantly different. These differences will be discussed in the following sections. However, the published dataset of CLASS and MPEX results is deeply unbalanced, meaning there are many combinations of teaching method, class size, and student population that have no data or very few data points in them (see Tables 3-5). This means that we are not fairly testing all combinations of factors and cannot accurately determine the influence of factors associated with the sparse or missing data. Because the data set is unbalanced and there are many cells with zero data (tables 3-5), it is important to avoid over-interpreting these results. Since we are drawing data from the published record, the unbalanced nature of the data is unavoidable.

**Table 2.** Results of three-way type 1 ANOVA comparing the effect of student population, class size, and teaching method on shifts on the CLASS and MPEX. All three factors were significant at the $p<.01$ level. The three- and two-way interactions were not significant. "df" = degrees of freedom.

| **Factor** | **df** | **F** | **p** |
|---|---|---|---|
| Student Population | 3 | 15.8 | $7.9 \times 10^{-7}$ |
| Class Size | 2 | 5.4 | .009 |
| Teaching Method | 3 | 27.4 | $1.3 \times 10^{-9}$ |

**Table 3.** Number of courses for each combination of student population and class size, " – " indicates no courses have a given combination.

| | | **Student Population** | | | |
|---|---|---|---|---|---|
| | | Algebra-based | Calculus-based | Elementary and Non-scientists | Upper-level |
| **Class Size** | Small | 1 | 9 | 16 | 2 |
| | Medium | - | 5 | 7 | 3 |
| | Large | 1 | 10 | - | 2 |

**Table 4.** Number of courses for each combination of teaching method and student population, "-" indicates no courses have a given combination.

| | | Teaching Method | | | |
|---|---|---|---|---|---|
| | | Focus on modeling | Explicit focus on developing beliefs | Some focus on developing beliefs | Ordinary courses (reformed and traditional) |
| **Student Population** | Algebra-based | - | 1 | - | 1 |
| | Calculus-based | 6 | - | 2 | 16 |
| | Upper-level | - | 1 | 2 | 4 |

**Table 5.** Number of courses for each combination of class size and teaching method, "-" indicates no courses have a given combination.

| | | Class Size | | |
|---|---|---|---|---|
| | | Small | Medium | Large |
| **Teaching Method** | Focus on modeling | 18 | 5 | - |
| | Explicit focus on developing beliefs | 1 | - | 1 |
| | Some focus on developing beliefs | 1 | 2 | 3 |
| | Ordinary courses (reformed and traditional) | 8 | 8 | 9 |

Below we describe how teaching method, class size, student population, and pretest scores may influence shifts on the CLASS and MPEX.

### A. Impact of teaching method on students' beliefs

The literature on the CLASS and MPEX is full of claims focusing on how different teaching methods or curricula lead to differing shifts in CLASS or MPEX scores. After one semester of traditional or research-based reformed physics instruction, it is common to report negative shifts on these surveys[1,2]. There are many other studies that find positive shifts in beliefs and contain claims about the teaching method that lead to these shifts (Table 6). There is very little mention of other factors besides teaching method or curriculum that may influence CLASS/MPEX shifts. In this section, we test some of these claims by aggregating the data for different types of teaching methods and comparing our results to claims in the literature.

**Table 6.** Teaching methods associated with positive shifts on the CLASS or MPEX as documented in the literature.

| |
|---|
| Courses that engage students in authentic scientific practices like building models of the physical world[26,27]. |
| Course for life science majors organized around rich biological models and taught using interactive engagement techniques[15] . |
| Explicit focus on epistemological development[4]. |
| Epistemological curriculum for life-science majors[29]. |
| Teaching physics in the context of its historical development[23]. |
| Physics and Everyday Thinking curriculum[22] with a special focus on learning about learning[31]. |
| Physics by Inquiry curriculum[21]. |
| No explicit instructional intervention[12]. |
| Modern physics course with reasoning development, model building, and connections to real world applications[19]. |
| Serving as a Learning Assistant[25]. |

Figure 2 shows shifts grouped by teaching method. Three studies reported on courses with an explicit focus on developing students' expert-like beliefs about physics and all resulted in positive shifts[4,23,29]. Examples of strategies include labs which helped students see that physics involves refining and reconciling intuitive ideas[4], activities where students reflected on their learning process, explicit epistemological framing of the course and use of associated vocabulary[4], modified Peer Instruction with discussions of intuitive answers to questions[4], "epistemologized" tutorials emphasizing the reconciliation of intuitive thinking and formal scientific thinking[29], and focus on the development of scientific ideas throughout history[23].

There were three studies where the instructors paid "some" attention to developing expert-like beliefs. These studies showed a small negative shift, no shift, or a small positive shift in beliefs. Examples of strategies included developing reasoning skills such as making inferences from observations and understanding why we believe scientific ideas[19] and helping students make personally meaningful connections to the content to increase interest[15]. (Another study cited their attention to student beliefs but did not explain how they did so[16]).

Several studies reported on courses that focused explicitly on developing models of the physical world. Commonly used curricula include Physics by Inquiry[34], Modeling Instruction[26], and Physics and Everyday Thinking[35] (also Physical Science and Everyday Thinking). Almost all of the courses taught with these curricula resulted in large positive shifts in CLASS and MPEX scores. A few courses using these curricula had small positive shifts or no shift, but no courses had negative shifts. These curricula are structured so that students work in small groups to perform experiments and gather evidence in order to build models of the physical world. They also participate in small group and whole classroom discourse to understand, validate, and refine these models, mirroring the way scientists create new knowledge. These curricula differ in the amount that they explicitly focus on learning about learning and all but one of the courses using these curricula show positive shifts in students' beliefs. There are other studies that report large positive shifts that also focus on models of the physical world, but do not use the curricula listed above[12,28,30].

There were also studies that described traditional or reformed courses where teaching methods designed to improve beliefs or explicitly develop models of the physical world were not used[1,2,4,12–15,19,20] (we refer to these as ordinary teaching methods). These courses may have included some elements of model building, but this was not the main focus of the course. These courses were associated with positive, zero, or negative shifts in beliefs.

To test the effect of these teaching methods on shifts in beliefs, we conducted a follow-up one-way ANOVA with teaching method as the independent variable and shift as the dependent variable (details of initial ANOVA discussed in introduction). We found a significant difference in shift in CLASS/MPEX score based on teaching method ($F(3, 52) = 698.2$, $p<.0001$). Using a Tukey HSD posthoc pairwise comparison test, we measured the differences between these four methods (Table 7). Modeling and explicit focus are not significantly different from each other ($p=.99$), and they are both better than teaching methods with "some" focus (modeling $p<.001$, explicit focus $p=.08$), which is in turn better than ordinary teaching methods (modeling $p<.0001$, explicit $p<.0001$, some focus $p=.08$).

**Table 7.** Results of Tukey-HSD posthoc pairwise contrasts comparing shifts for different teaching methods.

| | Teaching Method and Shifts in CLASS/MPEX scores | | | |
|---|---|---|---|---|
| | Focus on Modeling shift: 9.3% | Explicitly focus on developing beliefs shift: 8.5% | "Some" focus on developing beliefs shift: 0.7% | Ordinary methods shift: -3.7% |
| Focus on Modeling shift: 9.3% | - | p=.99 | **p<.001** | **p<.0001** |
| Explicitly focus on developing beliefs shift: 8.5% | - | - | **p=.08** | **p<.0001** |
| "Some" focus on developing beliefs shift: 0.7% | - | - | - | **p=.08** |
| Ordinary methods shift: -3.7% | - | - | - | - |

Overall, these studies suggest that courses that focus explicitly on developing students' beliefs about physics and those that are explicitly focused on building models of the physical world lead to the greatest positive shifts in student beliefs. These courses help students reflect on their learning, engage with their intuitive ideas and formal scientific thinking, and understand how scientific knowledge is created. Courses with "some" focus on developing beliefs have marginally greater shifts than ordinary courses, so even paying some attention to developing beliefs is of benefit to students.

Traditional courses and reformed courses where students have large gains on conceptual assessment have a range of shifts, but most result in large negative shifts in beliefs. Strong conceptual understanding does not automatically result in improved beliefs about physics. Attention to the process of learning science is likely necessary to improve beliefs, but not sufficient. Below we explore other factors in addition to teaching method that may be influencing students' shifts in beliefs.

### B. Impact of class size on students' beliefs

In Figure 2, we notice that many of the courses with positive shifts also tend to be small classes. It could be that positive shifts in beliefs are directly related to class size, so this factor should be further investigated. Perhaps the small class environment allows different kinds of interaction and discourse that help develop expert-like beliefs or the kinds of teaching methods that help students develop expert-like beliefs (Modeling, Problem-Based Inquiry, Physical Science and Everyday Thinking etc.) also happen to be those that work with a small class size.

To investigate the effect of class size, we plotted class size versus shift from pre- to posttest for small, medium, and large class sizes (Figure 3). We chose the cutoffs for medium and large classes using our intuition of the dynamics of different sized classes and also looked for natural cut points in the data. Small classes were those with 12 to 38 students, medium classes enrolled 42 to 100 students, and large classes enrolled 115 to 448 students. We found that the average shift for small classes (5.4) was larger than that of medium classes (.94), and large classes (-1.7).

The results of the three-way ANOVA (discussed in the introduction) revealed a significant effect of class size on shifts and no interactions between class size and other factors. We followed up with a one-way ANOVA with class size as the independent variable and shift as the dependent variable. We found a significant main effect for class size ($F(2,53)=5.4$, $p=.007$). Using a Tukey HSD posthoc pairwise comparison test, we measured the differences between these three class sizes (Table 8). Small classes had significantly greater shifts in CLASS/MPEX shifts than large classes ($p=.008$). Small and medium classes are not significantly different from each other ($p=.11$). Medium and large classes are also not significantly different from each other ($p=.56$).

**Table 8.** Results of Tukey-HSD posthoc pairwise contrasts comparing shifts for different class sizes. Small classes have statistically significant greater positive shifts than large classes.

| | Class size and shifts in CLASS/MPEX scores | | |
|---|---|---|---|
| | Small<br>shift: 5.4% | Medium<br>shift: 0.9% | Large<br>shift: -1.7% |
| Small<br>shift: 5.4% | - | p=.11 | **p=.008** |
| Medium<br>shift: 0.9% | - | - | p=.56 |
| Large<br>shift: -1.7% | - | - | - |

### C. Impact of student population on beliefs

In addition to class size being an important factor, the student population in each course might also be important. For example, many of the courses with large positive shifts on the CLASS/MPEX are courses taught to elementary education majors and many with large negative shifts are calculus-based physics courses. It may be that certain student populations have larger shifts in beliefs about physics than others. We plotted the student population types versus shift on the CLASS and MPEX (Figure 4). We find that with almost every population type, there is a range of shifts from positive to negative, with the exception of elementary education majors and non-science majors, where the majority of the courses show positive shifts.

The results of the three-way ANOVA (discussed in the introduction) revealed a significant effect of student population on shifts and no interactions between student population and other factors. We followed up with a one-way ANOVA with student population as the independent variable and shift as the dependent variable. We found a significant main effect for student population ($F(3,52)=5.8$, $p=.002$). Using a Tukey HSD posthoc pairwise comparison test, we measured the differences between these four student populations (Table 9). Courses for elementary education and non-science majors have greater shifts than calculus-based courses ($p=.006$). Upper-level courses have greater gains than courses for elementary education and non-science majors ($p=.007$). All other combinations of student populations were not significantly different indicating there are differences in CLASS/MPEX shifts based on student population.

**Table 9.** Significant results of statistical comparison between student populations using Tukey HSD posthoc contrasts. Courses for elementary education and non-science majors have significantly greater shifts than calculus-based or upper-level courses.

| | Student population and shifts in CLASS/MPEX scores | | | |
|---|---|---|---|---|
| | Elementary education & non-science majors shift: 6.7% | Calculus-based shift: 0.2% | Algebra-based shift: 1.6% | Upper-level shift: -2.8% |
| Elementary education & non-science majors shift: 6.7% | - | **p=.006** | p=.71 | **p=.007** |
| Calculus-based shift: 0.2% | - | - | p=.99 | p=.70 |
| Algebra-based shift: 1.6% | - | - | - | p=.83 |
| Upper-level shift: -2.8% | - | - | - | - |

### D. Impact of pretest scores on shifts in students' beliefs

It is also important to determine if and how pretest scores are related to the shifts. It could be that those who come into the class with more or less expert-like beliefs may make larger shifts in scores, regardless of how the course is taught, how large the enrollment, or the student population enrolled in the course.

We plotted pretest score versus shift for our dataset (Figures 5 and 6). We conducted a one-way ANOVA to test the effect of pre-test score on CLASS/MPEX shift. We found a significant main effect ($F(1,48)=731.1$, $p<.0001$) of pretest score. This ANOVA indicates that as CLASS/MPEX pretest scores increase, shifts on CLASS/MPEX decrease, but before drawing conclusions from this analysis, we need to consider underlying factors that may be influencing the relationship between pretest scores and shifts, e.g., differences in shifts and pretest scores by student population or teaching method. To investigate how these additional factors are related to pretest score and shift, we color-coded the dots in the scatter plots to correspond to different teaching methods (Figure 5) and student populations (Figure 6). Further, we used a two-way type 2 ANOVA to first test the effect of teaching method and pre-test score on shifts (We choose a type 2 ANOVA because the order of the factors teaching method and pretest is not inherently meaningful). We found that pretest was marginally significant ($F(1,45)=2.95$, $p=.09$) and teaching method was highly significant ($F(1,45)=36.4$, $p<.0001$). So, teaching method is the strongest predictor of shift, even when we take into account pretest score.

We find a lack of relationship between pretest scores and shifts when teaching method is taken into account, but teaching method can't influence pretest scores, since the pretest is taken before any substantial teaching takes place. A factor that could influence pretest scores is student population. In this dataset, the courses that focus on model building are primarily taught to elementary education majors and non-science majors while ordinary courses are usually calculus-based. We used a two-way type 2 ANOVA to test the effect of student population and pretest score on shifts. We found that pretest was marginally significant ($F(1,45)=3.7$, $p=.06$) and student population was also marginally significant ($F(3,45)=2.7$, $p=.06$). In this case, neither pretest score or student population are highly predictive of shifts on the CLASS/MPEX.

Overall, we find that teaching method is the strongest predictor of shifts on the CLASS/MPEX, even when pretest scores are taken into account. Student population and pretest score are marginally predictive of these shifts. These data suggest that teaching method is the most important factor for determining shifts in beliefs about physics, but pretest score and student population may also be important. This analysis would be more powerful if we had a balanced dataset (discussed in the Introduction). Further, if we could examine the data by individual student rather than data averaged over the entire class, we could better determine how the characteristics of students influence pretest score.

### E. Summary

Our meta-analysis of the factors that may influence CLASS/MPEX overall shifts is consistent with claims in the literature about the influence of teaching methods, but class size and student population may also be important factors. Teaching method explains the largest amount of variation in the CLASS/MPEX shifts, followed by student population, and then class size, with all three factors being statistically significant. This is consistent with our observation that the plot of shifts in beliefs grouped by teaching method (Figure 2) has the least variability within each category when compared to the plots of class size and student population (Figures 3 and 4).

We did not find any significant interactions between teaching method, student population, and class size, though we observe in the data that courses with large positive shifts are those with an explicit focus on model building, small class sizes, and taught to elementary education and non-science majors. The courses that tend to have negative shifts in beliefs are those taught with traditional or reformed teaching methods, large enrollments, calculus-based, and higher incoming beliefs. This lack of significant interactions between factors may be related to the large selection effects in this data set as a whole in terms of what kinds of classes researchers choose to study and report on and what factors they focus on.

We also found that teaching method is the strongest predictor of shifts on the CLASS/MPEX, even when pre-test scores are taken into account. Student population and pre-test score are marginally predictive of these shifts.

As mentioned above, the overall dataset is deeply unbalanced, so conclusions should be interpreted with caution. Researchers should fill in the gaps in this published record by focusing on factors beyond teaching method, so that we can determine how these other factors influence beliefs. For example, instructors should try to get large CLASS/MPEX gains in a large lecture class or an upper-division class using teaching methods that are successful in small introductory classes (though we acknowledge this will be difficult as many of these methods are designed for smaller courses). The CLASS/MPEX should be given to more classes for elementary education teachers that are taught using standard methods to determine whether the curricula with a focus on model building are leading to the large positive shifts, or if there is something unusual about this student population. The CLASS/MPEX should also be given in algebra-based courses, because there are only two published studies on this student population.

All of the data from the published record included in this meta-analysis are at the course level rather than the student level, giving us a coarse-grained view of students' beliefs. These data only indicate on average how teaching method, student population, class size, and pretest score influence shifts. Further, we have no information about within-class variation in shifts. If we instead used student-level data, we would be able to analyze the distribution in students' beliefs across the course and determine how the different factors investigated influenced this distribution. We could also determine how individual students' majors influence shifts instead of looking at the student population the course was intended for.

Our new Assessment Data Explorer, being developed as part of PhysPort[36], will address both of these concerns by collecting student-level assessment data. The assessment data explorer is discussed further in the "Future Directions" section below.

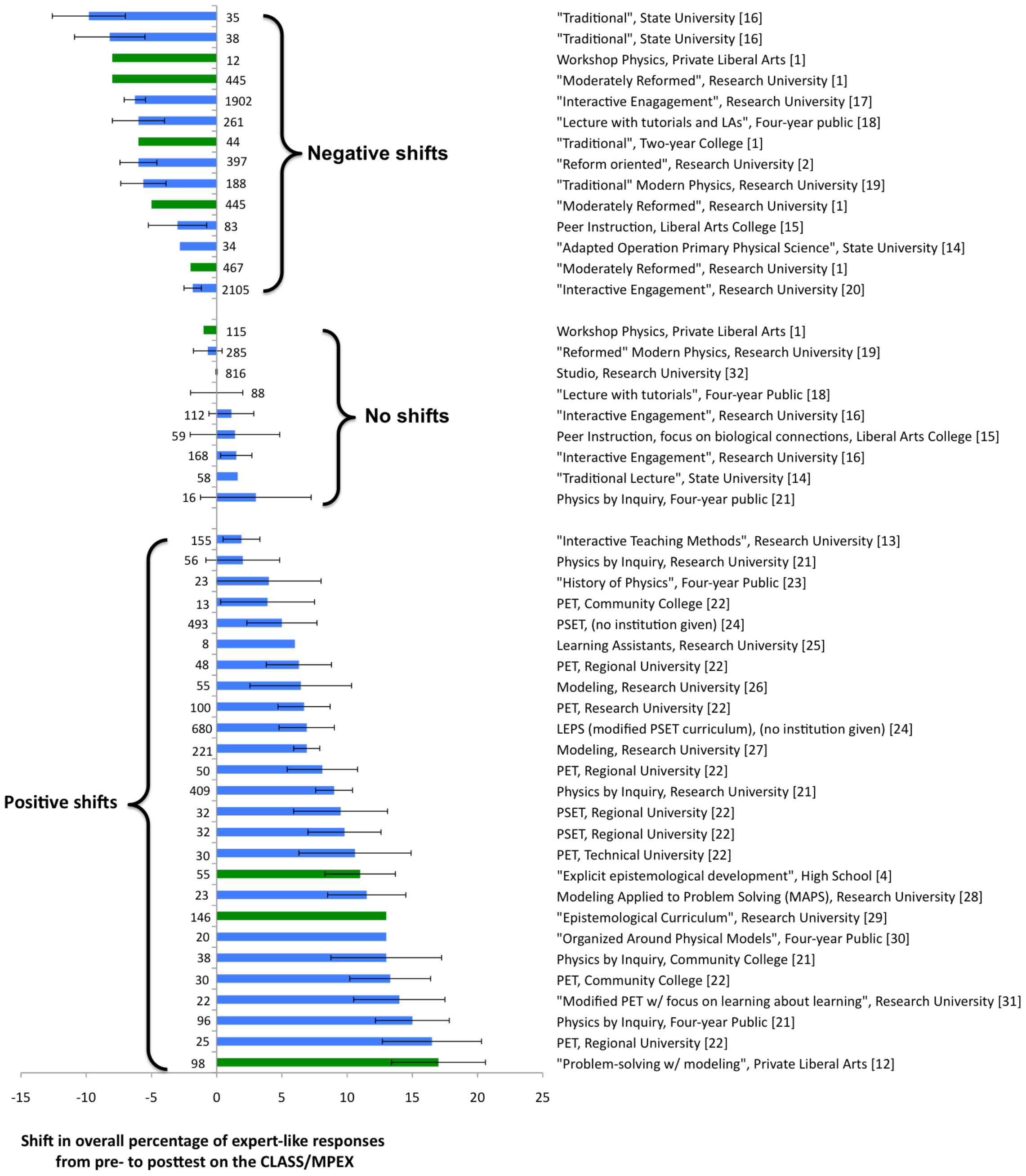

FIG 1. Shifts in percent expert-like on CLASS or MPEX scores from pre- to posttest based on direction of the shift. Numbers next to bars indicate the number of students in the study. Error bars represent standard error. Error bars were not available for all studies. Green bars indicate that the MPEX or MPEX-II was used. In all other cases the CLASS was used. PET = Physics of Everyday Thinking curriculum, PSET = Physical Science of Everyday Thinking curriculum.

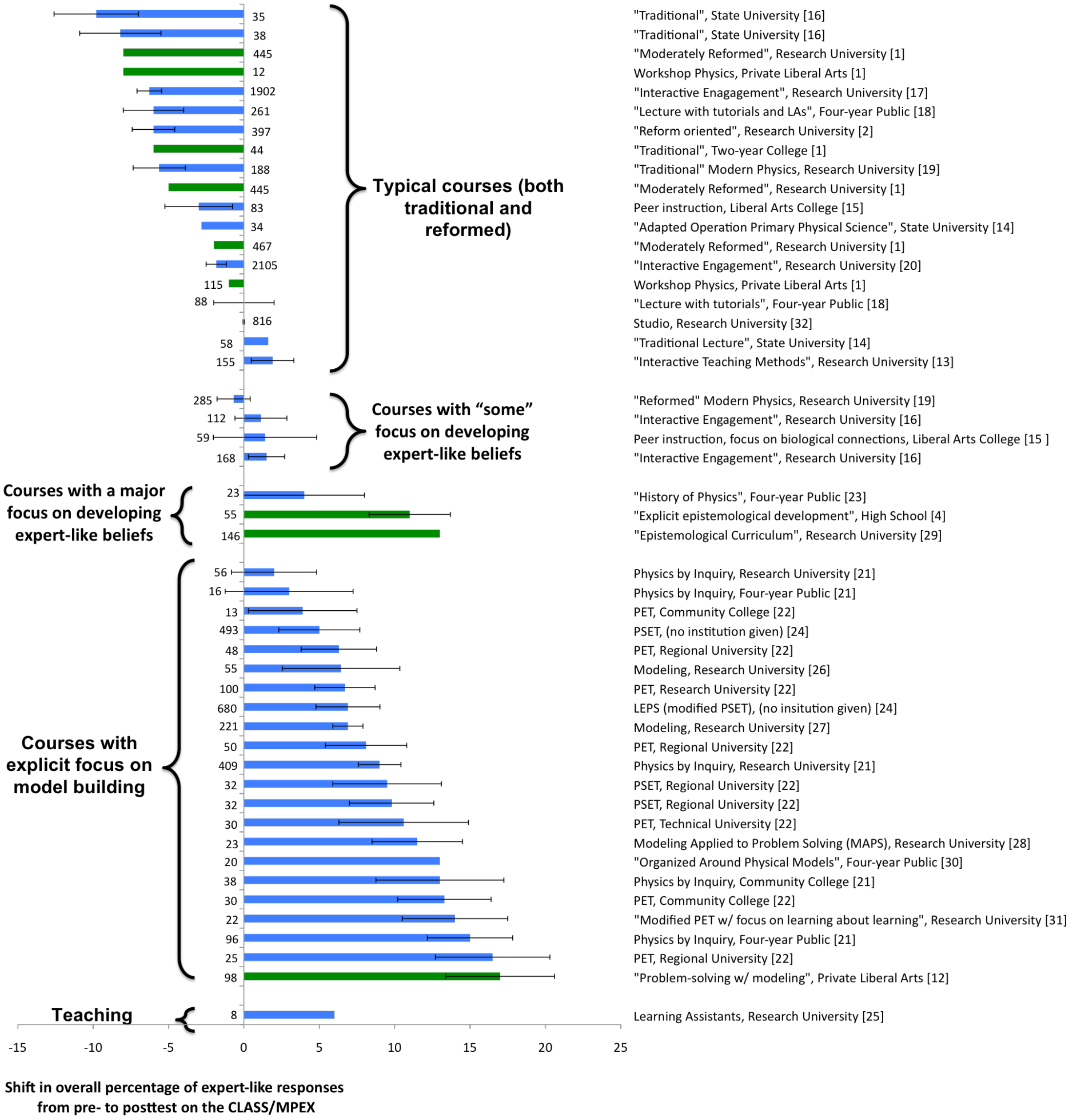

FIG 2**.** Shifts in percent expert-like on CLASS or MPEX by teaching method. Courses taught with traditional and reformed methods showed varying levels of positive and negative shifts. Courses that paid some attention to developing beliefs had no shift. Those that put a major focus on developing positive beliefs had positive shifts. Courses that explicitly focused on model building also had positive shifts. One small study of learning assistants found that these students also had a positive shift in beliefs. Classification of courses was determined by authors based on description of course provided in paper. PET = Physics of Everyday Thinking curriculum, PSET = Physical Science of Everyday Thinking curriculum.

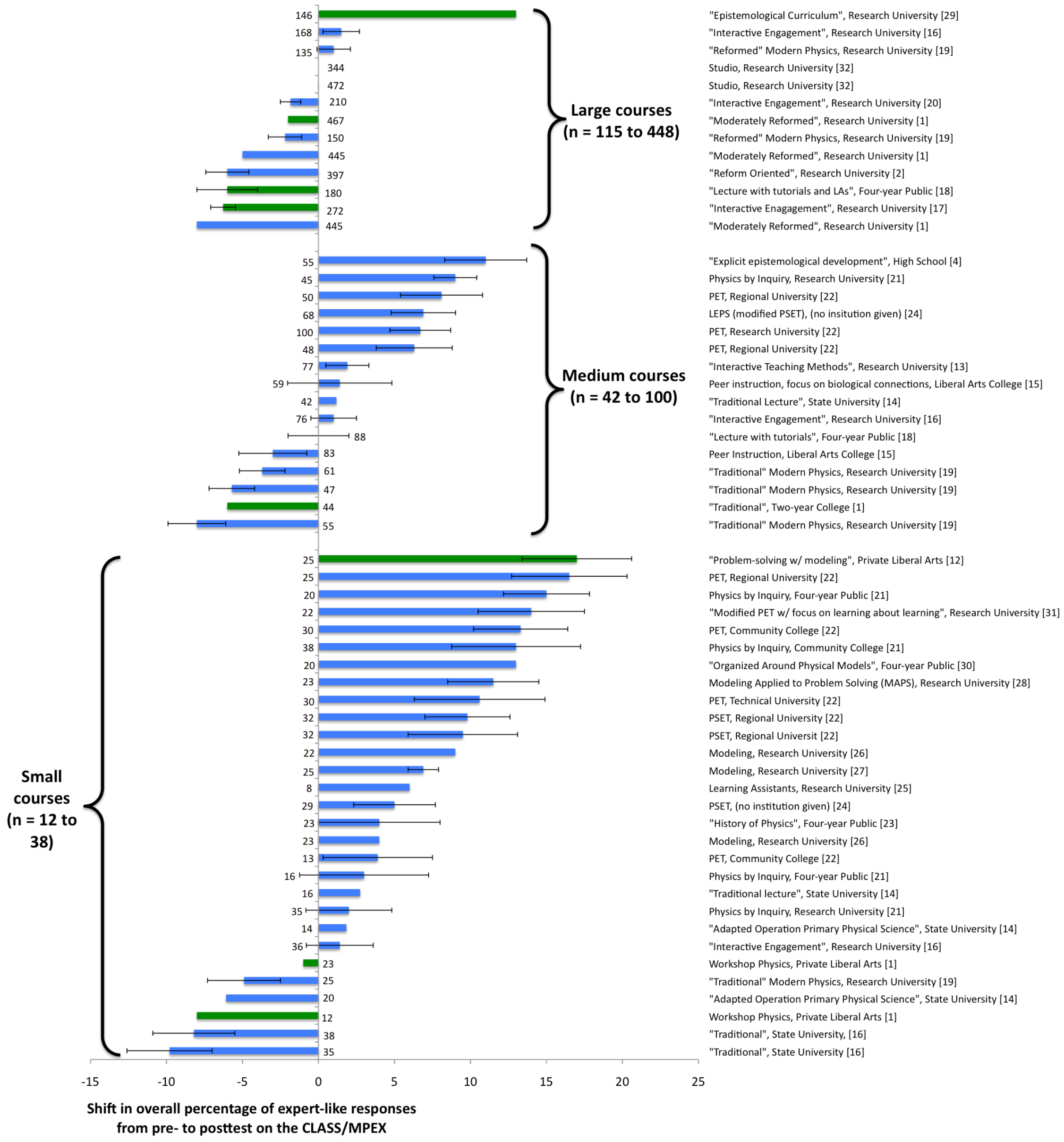

FIG 3. Shifts in CLASS or MPEX by class size. The average shift for small classes is positive and for large classes is negative. There is a bar for each unique course at a given institution taught with a given teaching method. In the case of the same course taught for multiple semesters, we used a weighted average to combine the results and display as a single bar. The total number of students is used as the class size for large lecture courses where students attended smaller lab or recitation sections. There were several studies where the total number of students and number of course sections were reported. Here we found the average class size by dividing the total number of students by the number of sections.

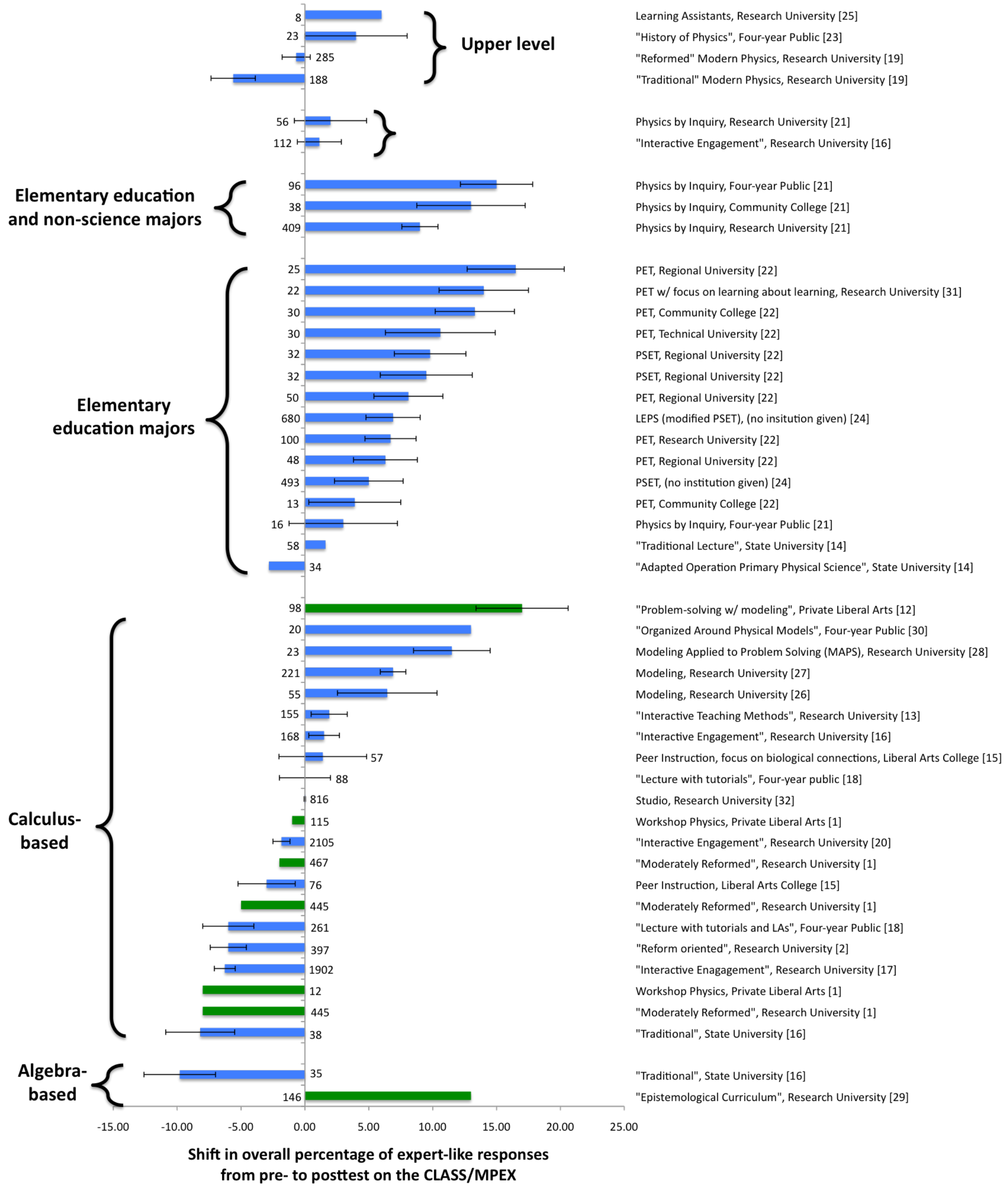

FIG 4. Shifts in CLASS or MPEX by student population. Nearly all student population categories include courses with both positive and negative shifts on the CLASS/MPEX, with the exception of courses for elementary education majors and courses for both elementary education majors and non-science majors, where all but one of the shifts are positive.

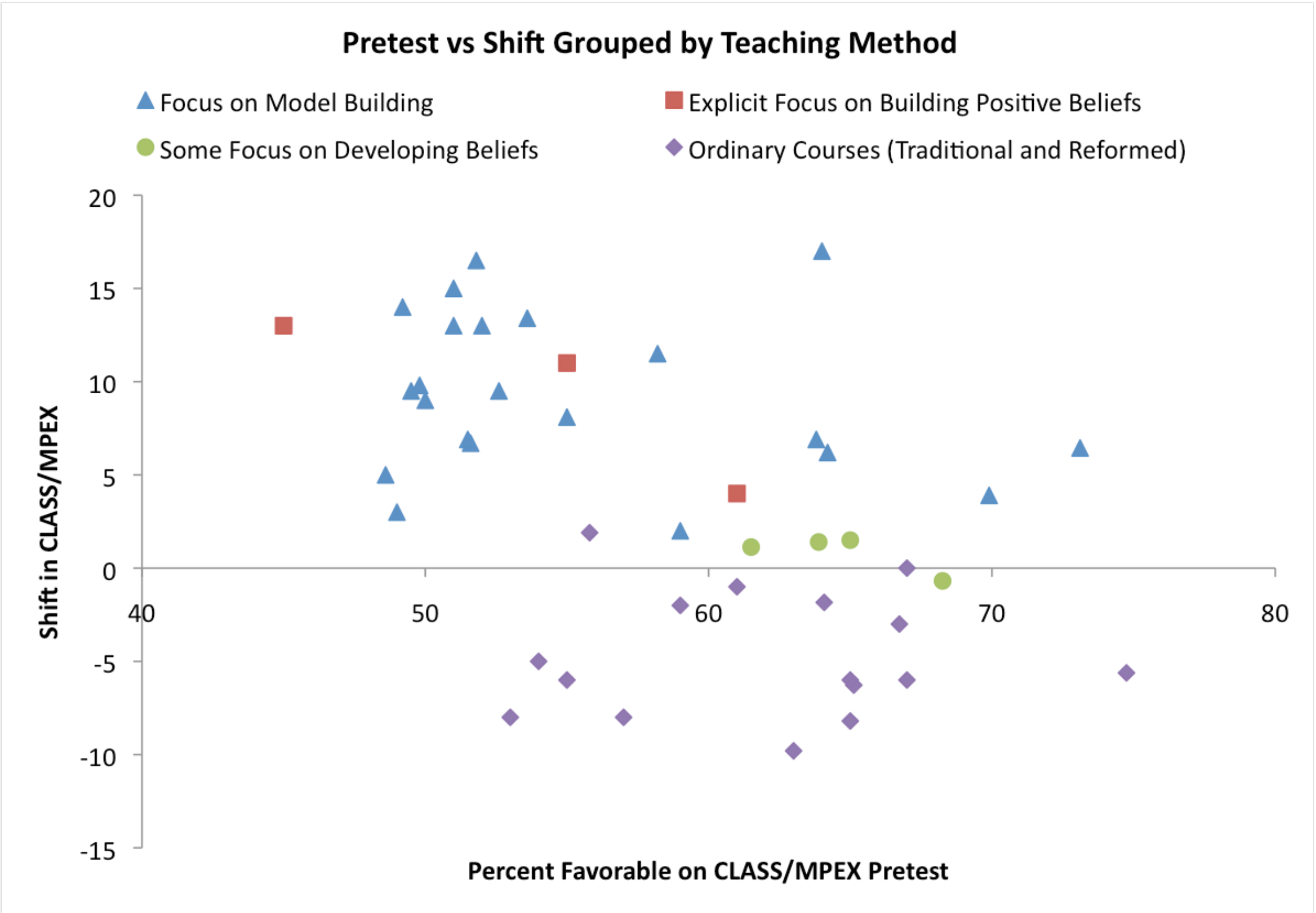

FIG 5. Pretest percent expert-like scores versus shifts on CLASS/MPEX with data points grouped by teaching method.

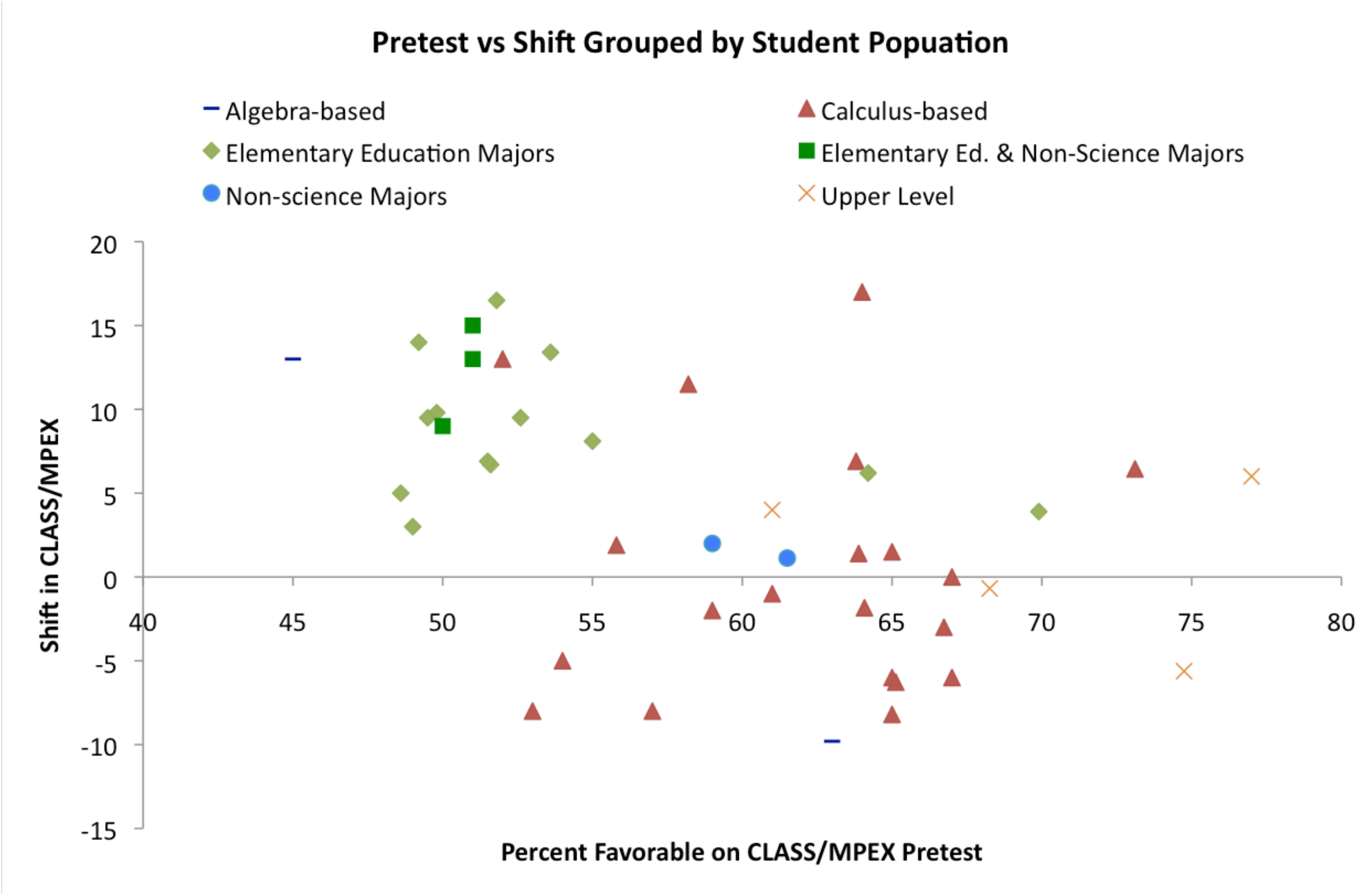

FIG 6. Pretest percent expert-like scores versus shifts on CLASS/MPEX with data points grouped by student population.

## IV. WHEN AND HOW DO PHYSICS MAJORS DEVELOP EXPERT-LIKE BELIEFS?

It has been shown that most advanced physics students have expert-like beliefs about learning physics[37,38] and also that most ordinary introductory physics courses result in negative shifts in beliefs (unless the instructor does something special), so how do these physics majors develop the expert-like beliefs that allow them to succeed in their major and go on to be successful physicists? It could be that they develop these beliefs in their undergraduate education or perhaps those who choose to major in physics already had positive beliefs about physics, presumably developed during their K-12 education.

Below we describe the results of several studies that examine beliefs of physics majors over time. These studies find physics majors start their undergraduate education with higher CLASS scores than other majors. Even those intending to major in physics in high school have higher CLASS scores than those not intending to major in physics. This indicates that physics majors develop these expert-like views in their K-12 education. These scores are relatively stable over the course of their undergraduate program when tracked longitudinally or looked at cross-sectionally. There is a jump in scores from the end of the undergraduate program to the beginning of post-graduate work, though small numbers make this conclusion tentative. The details of these studies are described below. All of these results suggest that physics courses are selecting students with expert-like beliefs, rather than developing them.

### A. Cross-sectional studies of development of beliefs

Cross-sectional studies of beliefs take snapshots of different students' beliefs in a given year, for example, collecting belief data from freshman, junior, sophomore, and senior physics majors at the same time in the same year and comparing them. These kinds of studies let us compare the beliefs of students at different points in their education and determine how their courses influence their beliefs over time. This kind of data is limited by the fact that many factors in addition to year in school could influence students' beliefs, for example, differing entrance requirements, class culture, class sizes, or teaching experience of faculty and the effect of these different factors is not easily differentiated.

Two cross-sectional studies found that physics majors' expert-like belief scores are relatively consistent across years. Bates et al.[39] surveyed physics majors in years 1-4 of their undergraduate curriculum, high school students, and a group of post-docs and faculty in a given year. They found no significant differences in the percentage of expert-like responses across years 1-4 of the undergraduate curriculum with the exception of a drop in scores in year 3. They believe this drop in year 3 data is anomalous. Bates et al. also compared scores of high school students intending to become physics majors and first year students intending to become physics majors and found no differences. They did find a statistically significant jump between year 4 students and post-graduates.

Gire et al.[38] compared students' beliefs in years 1-3 of the undergraduate curriculum in a given year and found no differences in scores by year. There were a small number of year 4 (n=16) and graduate students (n=7), so these are not included in their analysis, but the raw scores for year 4 and graduate students are higher than scores for years 1-3. Neither study controlled for dropouts, so any differences in scores between students in different years could be due to students with less expert-like physics not continuing in the major. That is, there could be a selection effect and scores in later years could increase because only students with expert-like beliefs about physics continued in their physics education.

### B. Longitudinal studies of development of beliefs

Longitudinal studies follow a cohort of students throughout their undergraduate physics program and compare their expert-like belief scores over time[38,40]. These studies allow us to draw stronger conclusions about how students' beliefs change over time than cross-sectional studies, but the data are more difficult to collect due to the longer timeframe and requirement to follow the same students. Two longitudinal studies of beliefs found that individual students' scores do not change over the course of their undergraduate program. Slaughter et al.[41] followed a cohort of 35 students through their first three years of the physics program. The CLASS was given as a pre- and posttest the first year, and a posttest in years 2 and 3. They found no statistically significant differences between years. Similarly, Gire et al.[38] found that most students' responses are stable over years 1-3 of the physics program, with 70% of students changing their answers by less than two questions (in this study, 51 physics majors responded to the CLASS more than once, but not all participants responded three times during years 1-3). These studies in tandem with the cross-sectional studies indicate that physics majors maintain their beliefs about learning physics over the course of the first three years. It is not their university courses that help them develop their beliefs; they already have them coming in.

### C. How does the development of beliefs of physics majors compare to other majors?

Several studies have compared the beliefs of physics majors and other majors over time and they all conclude that students who major in physics enter the physics program with more expert-like views than those who don't major in physics and these views are developed in their K-12 education. Perkins and Gratny[41] collected pretest CLASS data and intended major for students in their first university physics class. They waited several years and identified those students who actually majored in physics. They compared those who intended to major in physics, actually majored in physics, and majored in something else. They found that percent expert-like belief scores of those who actually majored in physics (78.3 ± 1.4%) were significantly higher than those who intended to major in physics (73.5 ± 1.2%) and the overall population (64.7 ± 0.3%). Gire and Jones[38] found that CLASS scores for physics majors in years 1-3 were higher than those of first year engineering students. Bates[39] found that CLASS scores of high school students intending to major in physics were similar to those of first year physics majors, but high school students not intending to major in physics scored lower than those intending to major in physics and first year physics majors.

### D. Conclusion

The cross-sectional studies, longitudinal studies, and comparisons between physics majors and other majors all point to the same conclusion: students who major in physics enter their university education with more expert-like beliefs than other majors and these beliefs are relatively stable over the course of their undergraduate career. This suggests that our undergraduate programs are not helping physics students develop more expert-like views of physics; they are only selecting for students who have developed these beliefs elsewhere. Further, the finding that students develop their expert-like views in their K-12 education raises the question, “what leads to the development of these expert-like beliefs?” These studies do not contain information about what kinds of physics instruction these students’ had in high school, so we can’t determine if these expert-like beliefs were developed as a result of teaching methods. We do know that most high school courses have small enrollments, and that students in small classes have significantly (or nearly significantly) more expert-like beliefs in our dataset of university students (See section IIIA). While small class sizes may contribute to the development of expert-like beliefs in students’ K-12 education, this does not explain why high school students planning to major in physics had more expert-like beliefs than those not intending physics as a major.

## V. HOW DO STUDENTS’ BELIEFS IMPACT THEIR LEARNING OF PHYSICS CONTENT AND VICE VERSA?

Physics faculty often have multiple learning goals for their students, including: learning conceptual and analytical physics content, thinking like physicists, and developing expert-like beliefs. To what extent are conceptual knowledge gains (such as those measured on the Force Concept Inventory (FCI)[6] and the Force and Motion Conceptual Evaluation (FMCE)[7], or similar tests) correlated with beliefs shifts on the CLASS or MPEX?

This question is difficult to answer because the data are mixed and unbalanced. Many studies report conceptual gains on various research-based assessments and varying percent expert-like shifts on the CLASS/MPEX. See references [2,16–19] for studies that report conceptual *gains* on research-based assessments and *negative shifts* on the CLASS/MPEX. Studies which report conceptual *gains* on research-based assessments and *no shifts* on the CLASS/MPEX include[16,18,19,42]. There are also studies which report conceptual *gains* on research-based assessments and *positive shifts* on the CLASS/MPEX[13,23,24,26,28–30]. Table 10 summarizes studies which correlate conceptual gains or losses with beliefs shifts.

As with all correlational studies, a correlation between these measures does not imply a causal relationship. The published record includes correlations between different combinations of pre and post beliefs and pre, post and gain in knowledge being calculated. These relationships can be suggestive of an important relationship between beliefs and conceptual knowledge, but without information on other mediating variables, we cannot make causal inferences.

**Table 10.** Correlations between surveys of beliefs about learning physics and other measures of learning. Numbers in brackets indicate the confidence interval which is significant if zero is not in the interval. *indicates a significant correlation.

| Measures Correlated | University and Teaching Method | N | Correlation Coefficient & Significance |
|---|---|---|---|
| **Incoming Beliefs** | | | |
| FCI pretest and CLASS pretest | Ryerson University, Modified peer instruction[13] | 155 | .294 ($p<.05$)* |
| FMCE gain (if pretest score < 60) and CLASS pretest | University of Colorado, Reformed Calc-based[16] | 256 | 0.21 (p<.001)* |
| FMCE gain and CLASS pretest | University of Colorado, Reformed Calc-based[43] | 337 | .20 |
| **Outgoing Beliefs** | | | |
| FCI posttest and MPEX posttest | Michigan State University, Traditional lecture[44] | 84 | .24 [0.04→0.42]* |
| FCI posttest and CLASS posttest | Ryerson University, Modified peer instruction[13] | 155 | .258 (p<.05)* |
| FCI gain and MPEX posttest | Loyola Marymount University, Interactive engagement[45] | 37 | .52 [0.24→0.72]* |
| FCI gain and MPEX posttest | Michigan State University, Traditional lecture[44] | 84 | .17 [-0.05→0.37] |
| FMCE gain (if pretest score < 60) and CLASS posttest | University of Colorado, Reformed Calc-based[16] | 256 | 0.26 (p<.001)* |
| Final exam and MPEX posttest | Michigan State University, Traditional lecture[44] | 97 | .27 |
| Course grade and MPEX posttest | | | weak |

### A. Correlations between pre or post conceptual knowledge and beliefs

A correlation between students' incoming beliefs and incoming conceptual knowledge tells us about how students' experiences prior to this course have influenced where they start. Only one study analyzes the connection between the FCI *pretest* scores and overall percent expert-like *pretest* score on the CLASS. It found a small but significant positive correlation between these measures[13]. This study suggests that students with more expert-like incoming beliefs also have more conceptual knowledge. Presumably students with higher incoming conceptual knowledge have previously taken a physics course. Most ordinary physics courses result in negative shifts in beliefs, so it is perplexing that students who have likely taken a previous course have higher incoming beliefs. However, this correlation could be explained by the selection effects discussed in the previous section.

Another way to determine how students' beliefs about learning physics and conceptual knowledge are related is to look at students' outgoing beliefs and conceptual knowledge. A correlation between students' outgoing beliefs and outgoing conceptual knowledge (posttest scores) tells us how the conceptual knowledge students ended the course with is related to their beliefs at the end of the course. Since this is a relationship between measures taken at the end of the course, it is likely that specifics of the course influence it. Two studies[13,44] have calculated the correlation between CLASS or MPEX *posttest* percent expert-like scores and FCI/FMCE *posttest* scores (see Table 10 for more details). Both found a small positive significant correlation between these measures indicating that the conceptual knowledge students leave the course with is positively related to their beliefs about learning physics at the end of the course. The causal inferences we can draw from this correlation are limited, because these data don't tell us if the students with greater conceptual knowledge and beliefs also entered the course this way and how the course influenced their beliefs and knowledge.

### B. Correlations between gains on conceptual surveys and beliefs

To get closer to answering the question, “Do students’ beliefs about learning physics influence what they actually learn?” we can examine the relationship between students’ *incoming beliefs* and the conceptual knowledge *gained* from the course. This relationship suggests how students’ initial beliefs about learning physics may influence what they actually learn. It is plausible that those who have more expert-like beliefs about learning physics learn it more effectively because they are applying strategies that actually work. For example, students who believe that learning physics is all about memorization are likely to attempt to memorize facts about physics instead of constructing their own understanding of the set of rich interrelated concepts.

There are two studies that correlate normalized gain on the FCI/FMCE and students’ incoming beliefs. One study found a small statistically significant correlation[16] and the other found a small correlation but the significance was not tested[43]. Both of these studies were at the same university with calculus-based physics students. These studies suggest that students who start physics with more expert-like beliefs gain more conceptual knowledge. If this were true, it would imply that helping students develop expert-like beliefs about learning physics would contribute to their ability to learn physics concepts. But these studies alone are not enough to make any strong causal inferences about how initial beliefs influence learning.

We can also examine how students’ beliefs at the *end* of the course (posttest beliefs) are related to their *gain* in conceptual knowledge over the semester. This would tell us how what they learned in the course is related to what they believe about learning physics at the end of their course, though, once again, there are many things that happened in the course that may have influenced each of these. Three studies have looked at this relationship. One found a statistically significant positive correlation of a moderate size between FCI *gain* and MPEX *posttest* percent expert-like score[45] and another study found a small non-significant correlation for the same measures[44]. A third study found a small significant correlation between FMCE *gain* and CLASS *posttest*[16]. Together, these studies indicate the correlation between normalized gains on these mechanics conceptual tests and the surveys of beliefs about physics are small but present.

Most studies of reformed-based teaching methods result in substantially higher conceptual gains than traditional teaching methods. However, reform-based courses often result in large negative shifts in beliefs. This implies that the correlation between learning gain and shift is in some cases negative, where strong conceptual understanding does not automatically result in improved beliefs about physics.

### C. Conclusion

More work is needed before we can answer the question, “How are students’ beliefs about physics related to their learning of physics?” Studies find a small significant correlation between students’ incoming beliefs about learning physics and their conceptual gains in the course, but there are many other variables, specifically those describing who the students are and what the course is like, that should be examined before we draw conclusions about how incoming beliefs influence learning. This finding suggests that those who have more expert-like ideas about learning physics learn more physics. If this was true, it implies that even if all you care about is student learning, helping cultivate expert-like beliefs in your students could improve their ability to learn physics. However, studies which investigate these effects are limited and more work needs to be done. For example, the relationship between how students' conceptual knowledge changed from the beginning to the end of the course (gain) and their change in beliefs from beginning to the end of the course (shifts) has not been studied. This correlation would help us see how changes in conceptual knowledge are related to changes in beliefs.

Additionally, the link between beliefs shifts and conceptual gains should be studied in more depth. Studies need to consider other variables that may also be correlated to conceptual gains and initial beliefs, specifically variables describing who the students are, e.g., physics background, major, math background, and variables that describe what the course is like, e.g., type of instruction.

## VI. CONCLUSION AND DISCUSSION

Studies of beliefs about learning physics using the CLASS and MPEX indicate the teaching method is the most important factor that influences the shift in beliefs from pre- to posttest, but that class size and student population also explain significant portions of the variance in shifts (though less so than teaching method) This is consistent with claims in the literature about how teaching methods influence beliefs (see Table 6). We find that courses with an explicit focus on modeling or developing students' expert-like beliefs have significantly greater shifts in CLASS/MPEX scores than courses with "some" focus on developing expert-like beliefs or ordinary courses. We also find that small classes have significantly greater shifts than large classes. Further, courses for elementary education and non-science have greater shifts than calculus-based and upper-level courses.

We did not find significant interactions between teaching method, class size, or student population when testing their influence on shifts, but this may be a result of the unbalanced nature of the published dataset. In our dataset, most studies on small classes use teaching methods which focus on model building and are taught to elementary education majors where most studies which focus on courses taught with traditional or reformed teaching methods are in large classes taught to calculus-based students.

We also tested the influence of pretest score (incoming beliefs about learning physics) on shift, taking into account teaching method and student population. We found that teaching method is the strongest predictor of shifts on the CLASS/MPEX, even when pretest scores are included in the analysis. Pretest score and student population were found to be marginally predictive of these shifts.

To better answer the question, "Are the improvements in beliefs reported in the literature supported by teaching interventions, small class sizes, student population, pretest scores, or some combination of these?", researchers should focus on factors beyond teaching method. For example, instructors should try to get large positive CLASS/MPEX shifts in large lecture classes or upper-division classes using teaching approaches that are successful in small introductory classes. The CLASS/MPEX should be given to more classes for elementary education teachers that are taught using standard methods to determine whether the curricula with a focus on modeling building are leading to the large positive shifts, or if there is something unusual about this student population. The CLASS/MPEX should also be given in algebra-based courses, as there are only two published studies on this student population. Further, student-level data should be collected and reported so that we can understand within-class variation in shifts.

These studies also present some concerning findings: for most of our large ordinary calculus-based courses, students' beliefs get worse over the course of the semester. We expect students would better understand the discipline of physics and how to learn physics *after* completing a physics course. Instead, their beliefs become less expert-like and students leave their course believing that physics is, for example, about memorizing facts, plugging numbers into equations, and not relevant to their life. Many faculty hope that as a result of their taking a physics class, students will come to appreciate physics as a coherent and logical method of understanding the world and to recognize that they can use reason and experimentation to learn about the world, although this is not the case in many of our large courses. Tremendous progress has been made over the last 30 years to help students develop strong conceptual understanding in these courses. More work needs to be done to figure out how to support students in improving their beliefs about learning physics in these learning environments.

The small, pre-service teacher courses that focus on model building consistently result in positive shifts in beliefs. These curricula are structured so that students work in small groups to perform experiments and gather evidence in order to build models of the physical world. They also participate in small group and whole classroom discourse to understand, validate, and refine these models, mirroring the way scientists create new knowledge. Researchers should try to implement teaching environments similar to these in larger calculus-based courses and look at the effect on students' beliefs, though we acknowledge this will be difficult as these methods are designed for smaller courses.

Explicitly focusing on developing students' beliefs about learning physics can also lead to positive shifts. There are a wide variety of strategies to do this including using labs to help students view physics as refining and reconciling intuitive ideas[4], activities where students reflect on their learning process[4], explicit epistemological framing of the course[4], modified Peer Instruction with discussions of intuitive answers to questions[4], "epistemologized" tutorials emphasizing the reconciliation of intuitive thinking and formal scientific thinking[29], and focus on the development of scientific ideas throughout history[23]. Strategies such as these should also be studied with a wider variety of courses and student populations.

Physics majors' beliefs remain relatively unchanged over the course of the physics major and those who major in physics have more expert-like beliefs than other majors. This indicates that we are not helping physics students develop expert-like beliefs but instead those who become physics majors already have these expert-like views about learning physics. This is concerning, as those who enter the university with less expert-like beliefs aren't majoring in physics because this choice may be less readily available to them and they are less inspired to do so. A stronger focus on developing expert-like beliefs in introductory courses might allow a wider variety of students the opportunity to major in physics, though this finding is suggestive but weak. Future studies looking at how individual students' beliefs develop in their K-12 education, how these change in introductory classes, and the relationship between beliefs and choice of major would clarify this finding.

Students' beliefs about physics are weakly but significantly correlated with measures of their conceptual understanding. Students who have more expert-like incoming or outgoing beliefs also have greater gains on research-based assessments of conceptual understanding, such as the FCI and the FMCE. Specifically, there is a small correlation between students' incoming beliefs about physics and their normalized gain on these mechanics concept inventories. This suggests that students with more expert-like incoming beliefs gain more conceptual knowledge in their physics course. There are only two studies looking at this correlation, so the results are somewhat inconclusive. Further, there are many other important variables that could influence beliefs and conceptual understanding that have not been studied. We encourage researchers to further study the relationship between beliefs about learning physics and conceptual understanding and variables that may influence this relationship such as physics and math background and details of the course in order to better understand how each develops and provide clues for instructors on how to support this process.

## VII. FUTURE DIRECTIONS

Through this analysis of existing CLASS and MPEX data, we have identified several open questions about how students' shifts in beliefs relate to what happens in the classroom. To more definitively answer these questions, a larger data set that included individual student level data is needed. We are currently developing a national database of research-based assessment results (including the CLASS and MPEX) on PhysPort[36] (formerly the PER User's Guide). Here instructors can upload their student's de-identified assessment data in order to visualize and analyze their results in a variety of ways. A database of this kind will give the PER community access to the kind of data needed to answer open questions such as, "Are the improvements in beliefs reported in the literature supported by teaching interventions, small class sizes, student population, pretest scores or some combination of these?", "How strongly do beliefs depend on the population of students and their backgrounds?", and "Do expert-like beliefs support student learning of physics, is it the other way around, or are there other important variables that influence one or both that haven't yet been studied?" This database will be ready for use by verified physics instructors in the fall of 2015.

**Appendix**

**Table 11.** Comparison of the CLASS and MPEX surveys.

| | MPEX | CLASS |
|---|---|---|
| **Structure of survey** | 34 items<br>20-30 minutes to complete | 42 items[1]<br>8-10 minutes to complete |
| | 5 point Likert scale (strongly agree, agree, neutral, disagree, strongly disagree) | |
| **Focus of survey** | Student beliefs about the physics course | Student beliefs about the discipline of physics |
| **Typical administration** | Pre- and posttest (beginning and end of course)<br>Either online or paper and pencil<br>In class or at home | |
| **Survey development process** | Tested over the course of four years at 15 universities with over 1500 students. Items chosen through literature review, discussion with faculty and the researchers' personal experiences. | Used seven design principles (see[2]) to revise MPEX and VASS questions and create new ones. Conducted interviews with students and relied on the language and ideas students used to revise and create new questions. |
| **Validation of survey items** | Student interviews where students were asked to explain why they chose each answer. | |
| **Student population tested** | Primarily calculus-based physics students | Conceptual physics, algebra-based and calculus-based physics students |
| **Source of "expert" or "favorable" response** | 19 physics faculty implementing Workshop Physics. 80% agreement level reached on all but three survey items. | 16 physics education researchers or physicists involved in teaching. 100% agreement level on all but four survey items. |
| **Scoring** | To find the *percent of expert-like response* for each student, the number of "agree" and "strongly agree" responses are added together and divided by the total number of survey items. A similar process followed to find the *percent of novice-like responses*.[2] | |
| **Categories** | • Independence<br>• Coherence<br>• Concepts<br>• Reality link<br>• Math link | • Real-world connections<br>• Personal interest<br>• Sense-making/effort<br>• Conceptual connections<br>• Applied conceptual understanding |

[1] The CLASS has more items than the MPEX (42 vs. 33) but it takes less time for students to respond to them, presumably because the items on the CLASS are easier to parse.

[2] It has been found that students don't consistently interpret agree and strongly agree in the same way, so combining these results is appropriate[46].

| | | |
|---|---|---|
| | • Effort | • Problem solving general<br>• Problem solving confidence<br>• Problem solving sophistication<br>27 items are in one of these categories, 9 more are valid and reliable, 8 items are not scored because there is no expert response or the validity has not been established. |
| **Category creation** | Categories decided *a priori* by researchers - a deliberate decision consistent with the researchers' *resources* theoretical viewpoint in which students' beliefs are viewed as local coherences, not stable mental structures[46]. Statistical analyses found that survey items in some categories are only weakly correlated in actual student responses[2]. | Categories created using reduced-basis factor analysis, where raw statistical categories and categories predetermined by researchers were combined iteratively. They ensured that the questions in each category were correlated in actual student responses. |
| **Limitations** | Measure self-reported beliefs. | |
| | Several items contain two statements which are sometimes interpreted inconsistently by students. It is unclear how students who have never taken a physics course interpret the questions since they are strongly grounded in the physics course. | Test contains un-scored items for which validity has not been established. |